# In-Silico Investigation of 3D Quantitative Angiography for Internal Carotid Aneurysms Using Biplane Imaging and 3D Vascular Geometry Constraints

Kyle A. Williams,[a,d] SV Setlur Nagesh,[c,d] Daniel R. Bednarek,[a,b,d] Stephen Rudin,[a,b,c,d] Ciprian Ionita[a,b,c,d,*]

[a]University at Buffalo, Department of Biomedical Engineering, Buffalo, USA

[b]University at Buffalo, Department of Radiology, Buffalo, USA

[c]University at Buffalo, Department of Neurosurgery, Buffalo, USA

[d]Canon Stroke and Vascular Research Center, Buffalo, USA

**Background:** Quantitative angiography (QA) in two dimensions has been instrumental in assessing neurovascular contrast flow patterns, aiding disease severity evaluation and treatment outcome prediction using data-driven models. However, QA requires high temporal and spatial resolution, restricting its use to digital subtraction angiography (DSA).

**Purpose:** The 2D projective nature of DSA introduces errors and noise due to the inherently three-dimensional flow dynamics. This study examines whether 3D QA information can be recovered by reconstructing four-dimensional (4D) angiography using data from standard clinical imaging protocols.

**Methods:** Patient-specific 3D vascular geometries were used to generate high-fidelity computational fluid dynamics (CFD) simulations of contrast flow in internal carotid aneurysms. The resulting 4D angiograms, representing ground truth, were used to simulate biplane DSA under clinical imaging protocols, including projection spacing and injection timing. 4D angiography was reconstructed from two views using back-projection constrained by an a priori 3D geometry. Quantitative angiographic parametric imaging (API) metrics obtained from the CFD-based 4D angiography and reconstructed 4D angiography, respectively, were compared using mean square error (MSE) and mean absolute percentage error (MAPE).

**Results:** The reconstructed 4D datasets effectively captured 3D flow dynamics, achieving an average MSE of 0.007 across models and flow conditions. API metrics such as PH and AUC closely matched the CFD ground truth, with temporal metrics showing some variability in regions with overlapping projections. These results demonstrate the potential to recover 3D QA information using simulated 4D angiography constrained by standard clinical imaging parameters.

**Conclusions:** This study highlights the feasibility of recovering 3D QA information from reconstructed 4D DSA simulated from biplane projections. The method provides a robust framework for evaluating and improving QA in clinical neurovascular applications, offering new insights into the dynamics of aneurysmal contrast flow.

**Keywords**: 4D angiography, 1000 fps, reconstruction, patient-specific phantom.

*Ciprian Ionita, E-mail: cnionita@buffalo.edu

# 1 Introduction

Intracranial aneurysms and related neurovascular diseases in the Circle of Willis currently lack objective diagnostic tools during image-guided interventions.[1, 2] Endovascular treatments—such as coiling or stenting with flow diverters—rely on imaging to evaluate treatment outcomes and disease severity.[3-5] Digital subtraction angiography (DSA) remains the primary modality for these procedures.[6, 7] Although DSA provides high spatial and temporal resolution, its interpretation depends on operator experience, which may lead to variability in assessing treatment efficacy and disease progression.[8]

Quantitative angiography (QA) has emerged as a complementary approach to standard imaging, providing objective metrics that quantify contrast flow patterns strongly correlated with underlying hemodynamics.[4, 9] QA has demonstrated utility in analyzing neurovascular diseases and predicting treatment efficacy.[2] A significant advancement in this field is angiographic parametric imaging (API),[9] which extracts discrete hemodynamic parameters from time-density curves (TDCs) at every pixel in an angiographic sequence (Sec. 2.4). These parameters enable quantitative mapping of underlying hemodynamics, yielding further insights during aneurysm diagnosis and treatment prognosis.[9] Despite its promise, API's application is constrained by the two-dimensional (2D) nature of DSA, limiting its ability to resolve the inherent three-dimensional (3D) complexity of contrast flow phenomena.[6]

Errors and artifacts introduced by overlapping vascular structures and the lack of depth information in 2D dynamic imaging emphasize the need for advanced methodologies capable of reconstructing three-dimensional flow dynamics.[10] Recent advancements in computational modeling, particularly the use of computational fluid dynamics (CFD), have provided robust tools for generating high-fidelity, time-resolved datasets that can serve as ground truth for validating new imaging techniques.[11] By leveraging patient-specific vascular geometries, CFD simulations offer a detailed understanding of contrast flow dynamics, paving the way for innovative approaches to QA.[12]

Building on these developments, this study explores a constrained back-projection reconstruction of four-dimensional (4D) angiography—integrating 3D spatial data from CTA with high temporal resolution biplane DSA imaging.[13] This methodology addresses the limitations of 2D imaging while adhering to standard clinical protocols.[12] Using CFD-generated ground truth data,[11] we evaluate the accuracy, preservation of functional information and clinical feasibility of this approach, aiming to improve the analysis of aneurysmal hemodynamics and enhance decision-making in neurovascular interventions.[14]

## 2 Methods

### 2.1 Experimental Setup

Angiographic imaging data was generated *in silico* using CTA-derived patient-specific internal carotid artery aneurysm (ICA) phantoms within a computational fluid dynamics (CFD) solver (ANSYS Inc., Canonsburg PA). Simulated angiography was used to both precisely control the flow parameters of our angiographic data, and to ensure direct, quantitative comparison was possible between ground truth and reconstructed angiograms.

First, patient-specific ICA phantoms were imported into ICEM, a CFD mesh generation software. Inlet and outlet boundaries were defined for each model, and simulation parameters consistent with previously reported CFD-based simulated 4D angiography were applied.[15] These parameters included transient, laminar flow simulations solving the incompressible Navier-Stokes equations with a time step of 1 ms to resolve all unsteady motions in the flow. The fluid's density and viscosity values, approximating blood (1060 kg/m³, 0.0035 Pa-s) and contrast media (1190 kg/m³, 0.0063 Pa-s), were defined at physiological temperature (37°C). Using the passive scalar method, tracer fluid was labeled at the arterial inlet for 1 second to generate a 4D dynamic angiogram with a one second contrast injection. Temporal coverage captured inflow and outflow dynamics within the vessel. This methodology was repeated for three intracranial aneurysm models with inlet velocities of 25 cm/s, 35 cm/s, and 45 cm/s, resulting in nine 4D angiography simulations.

To convert the 4D simulations into 2D biplane, dynamic angiograms, a cone-beam projection geometry was simulated in ASTRA.[16] Frontal and lateral views were configured at a 90-degree angle to provide perpendicular projections. To test our ability to align biplane images (Sec. 2.2), the source-to-image distance (SID) was set at 105 cm for the frontal view and 110 cm for the lateral view. The source-to-object distance (SOD) varied between 75 cm in the frontal view and 80 cm in the lateral view to create magnification differences. A pixel pitch of 0.194 mm was used for both views to replicate typical C-arm imaging configurations.

*2.2 Data Alignment*

An important step in successful constrained back-projection is alignment of both biplane DSA views with a common 3D vascular mask. This first requires equalization of the representation of vascular structures in both biplane views via standardization of the pixel size in each imager. Though we cannot determine pixel size directly without a calibration object, it is possible to ensure that the pixel size of one image is the same as the other by estimating the relative magnification of the vessel in each image.

To accomplish this relative magnification estimation, we utilize the shared axis between the biplane views to compare transit of contrast in the axial direction, as described in our previous work.[17] Once the respective images are scaled to the same pixel pitch (which is known for each detector), remaining discrepancies in this measurement are due to relative magnification differences. Taking the ratio of these measurements yields a scaling factor that can be applied to the pixels of one image system to effectively equalize pixel sizing (Eq. 1).

$$p'_{new} = \frac{l'}{l} p' \qquad (1)$$

Where p' is the pixel pitch of the primed detector matrix, and l and l' are the measured advancement of contrast between two frames from the unprimed and primed image systems, respectively. If needed, the scaling factor could be estimated from a collection of such comparisons, to improve the robustness of the estimation.

Since single pixels represent the same physical distance in each imager after rescaling, the only remaining discrepancy between the two systems is the FOV. We use another previously developed algorithm to equalize the axial fields of view (FOVs) of both imagers, which ensures the axial position of the vessel is represented by the same coordinates in both systems.[17] The distance between the highest axial point of contrast and the bottom of the image frame is compared and equalized between views at corresponding time points. To ensure all axial points in the image are matched, whichever imager has a smaller FOV is used as the reference, and the other image is cropped accordingly.

Once biplane imagers are aligned, the 3D volume of the vascular structures of interest must be incorporated. This can either be obtained directly from the angiography data via epipolar reconstruction [17-19] or from a previously acquired CTA. In the latter case, the volume will need to be registered to both biplane views prior to 4D reconstruction. To test this registration robustly, a cone beam CT (CBCT) acquisition, and subsequent volumetric reconstruction, was simulated for each model using ASTRA. A binary mask of the vasculature is generated from this volumetric reconstruction. Subsequent rigid registration of this 3D mask to 2D masks from both biplane views is facilitated by simpleElastix[20, 21], following a previously developed protocol for 2D to 3D image registration.[22, 23]

*2.3 4D Angiographic Reconstruction*

After geometric correlation of contrast flow information from both views and the known 3D vascular structure, it is possible to estimate a 4D angiogram, such that the 3D flow profile originally captured in each projection view is reconstructed into a series of time-resolved 3D volumes.

Using the techniques described in Section 2.2 to ensure both projections are geometrically equivalent, we first back-project the contrast intensities in each image into a common volume. We then multiply this back-projection matrix by the registered 3D vascular structure mask, spatially correcting the contrast profiles and improving localization of the contrast data. Additionally, intensity profiles are path-length corrected using the 3D mask during this step, which has demonstrated an ability to better preserve true contrast densities in previous studies.[15] This constrained back-projection technique is applied across the entire volume, and repeated for each time step, yielding a 4D angiogram, demonstrated in Figure 1. To determine how closely this reconstructed angiogram aligns with ground truth 4D CFD data, the two datasets were compared via mean squared error (MSE).

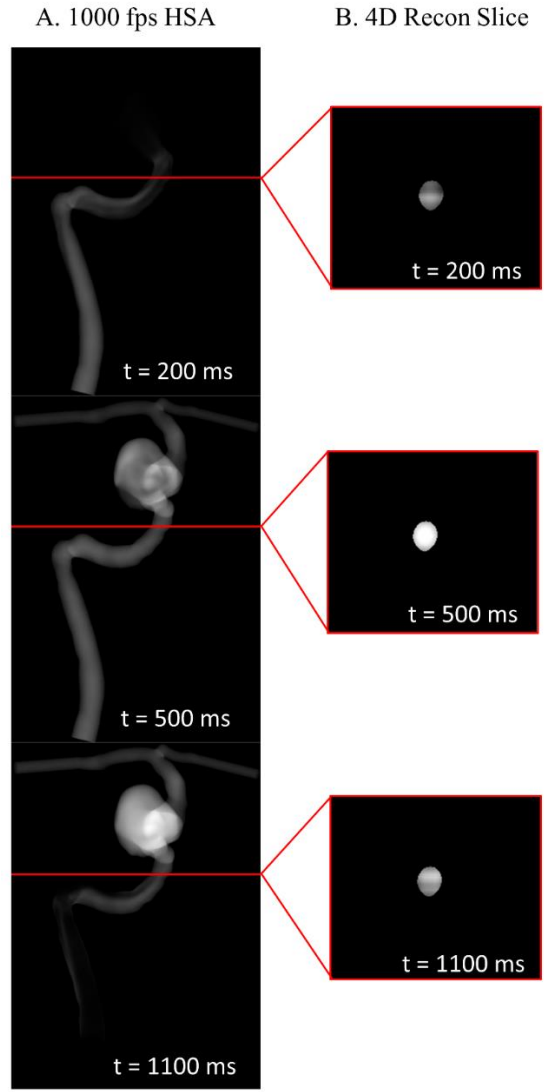

**Figure 1:** Constrained back-projection of the same axial slice at three different time points. Each row represents a different time point. The red line in the AP view of the original angiographic acquisition (A) is used to reference a specific slice of the 4D reconstruction (B) at the indicated time.

## 2.4 Angiographic Parametric Imaging (API)

Angiographic parametric imaging (API) is a form of quantitative angiography which parameterizes pixelwise time-density curves (TDCs) and reports metrics such as time-to-peak (TTP), mean transit time (MTT), time-to-arrival (TTA), peak height (PH) and area under the time-density curve (AUC), shown in Figure 2. Though these metrics are typically calculated for each pixel in projection angiography data, it is also possible to generate voxel-based TDCs, allowing this analysis to be translated to volumetric flow patterns. To assess the functionality of our reconstructed 4D angiograms to provide quantitative information, we perform an API analysis on both the reconstructed 4D data, and the ground truth 4D CFD-generated data. These resultant volumetric API maps are compared across datasets with such metrics as mean absolute error (MAE), MSE and mean absolute percentage error (MAPE).

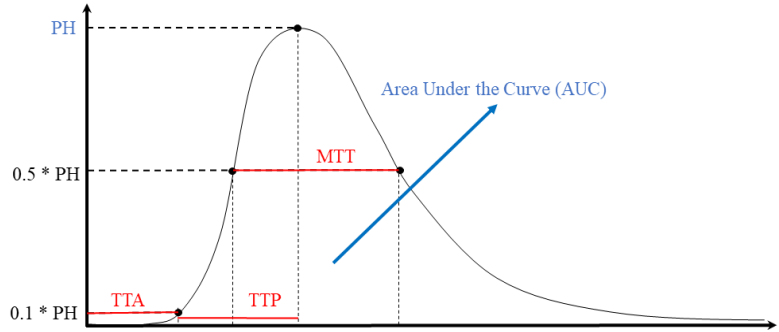

**Figure 2:** A sample time-density curve, with each API parameter denoted, including time-to-peak (TTP), mean transit time (MTT), time-to-arrival (TTA), peak height (PH) and area under the curve (AUC). Temporal parameters are shown in red, while intensity-based parameters are shown in blue.

## 3 Results

### 3.1 4D Angiographic Reconstruction

For each model, several volumes throughout the time series are displayed in Figure 3, which shows the inflow of contrast through each model. To represent the 3D structure effectively in a 2D figure, the volume is shown from several different angles throughout the sequence. The MSE between reconstructed volumes and ground truth 4D CFD data, averaged across all time-steps and across all velocities, is 0.013, 0.004 and 0.0043 for M1, M2 and M3, respectively. Table 1 shows the MSE as a function of velocity for each of these models.

**Table 1:** Quantitative comparison between reconstructed 4D angiograms and ground truth CFD simulations. Each MSE measurement represents the error across all time steps for a specific inlet velocity.

| Model | Inlet Velocity (cm/s) | MSE |
|---|---|---|
| M1 | 25 | 0.016 |
|  | 35 | 0.013 |
|  | 45 | 0.01 |
| M2 | 25 | 0.005 |
|  | 35 | 0.004 |
|  | 45 | 0.003 |
| M3 | 25 | 0.006 |
|  | 35 | 0.004 |
|  | 45 | 0.003 |

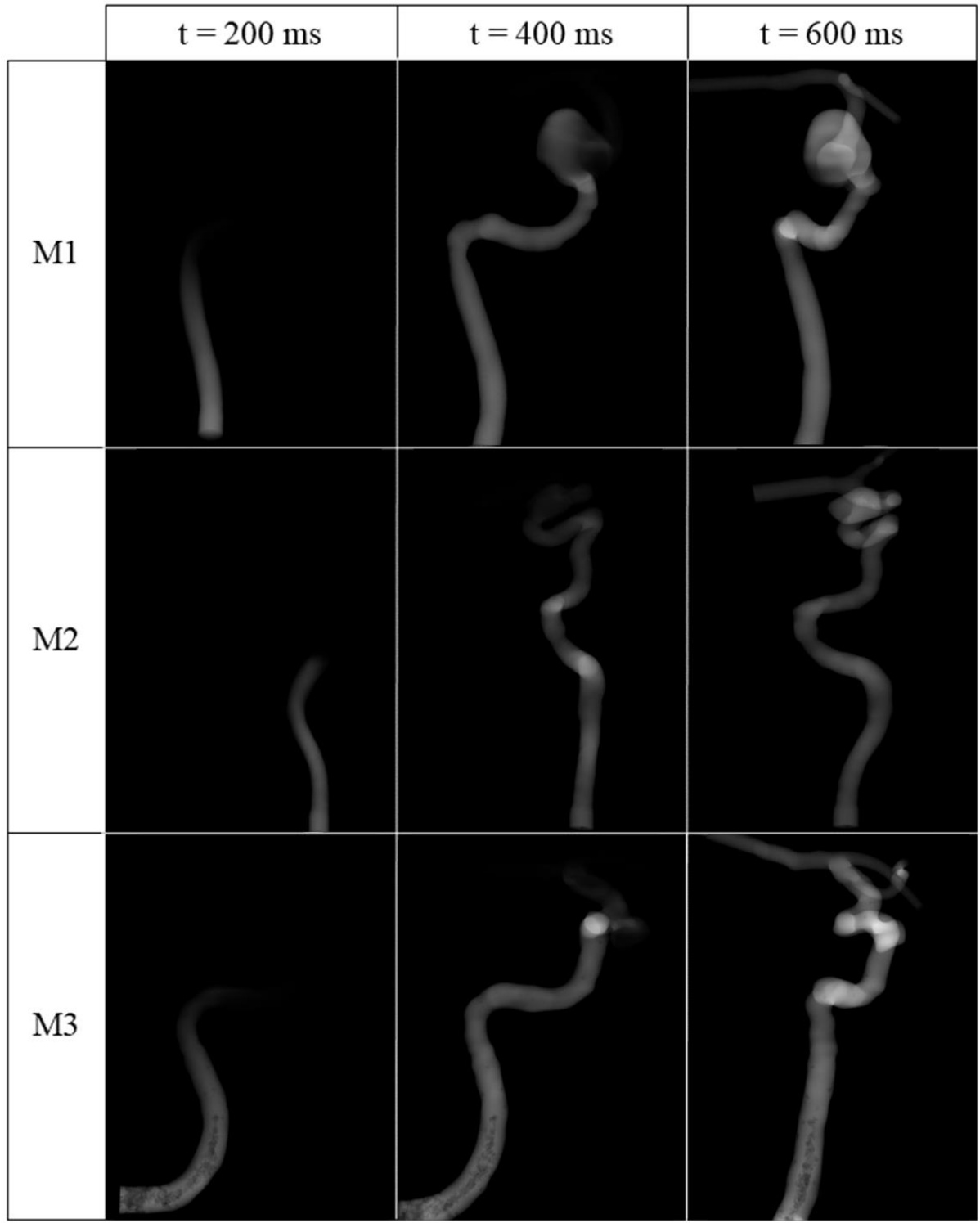

**Figure 3:** Representative time points from each model's reconstructed 4D angiogram are shown across each row. To demonstrate the volumetric nature of the angiograms, the time series is viewed from various angles throughout the sequence.

## 3.2 Angiographic Parametric Imaging (API) Analysis

API maps generated from 4D angiographic reconstructions of model M1 are displayed in Figure 4. The parameters are quantitatively compared to those generated from ground truth CFD simulations in Table 2. From the table, we observe strong agreement across all parameters, with the intensity-based parameters (PH and AUC) showing the best agreement. To further compare the volumes, Figures 5 and 6 show a cross-sectional comparison of API results within the inlet artery and aneurysm in M1 (respectively), generated from the two respective datasets.

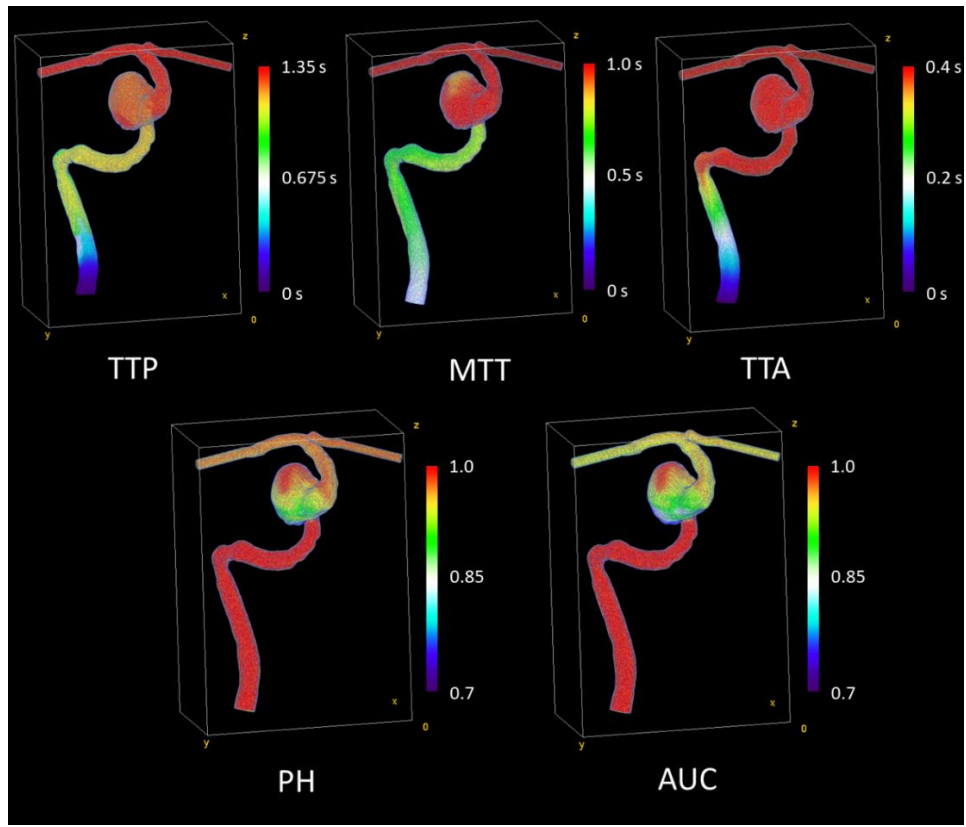

**Figure 4:** API maps for model M1 at 25 cm/s inlet velocity, generated from the constrained back-projection-reconstructed 4D angiogram. Maps include time-to-arrival (TTA), mean transit time (MTT), time-to-peak (TTP), peak height (PH), and area under the time-density curve (AUC). Temporal parameters are in seconds, while intensity-based parameters are unitless.

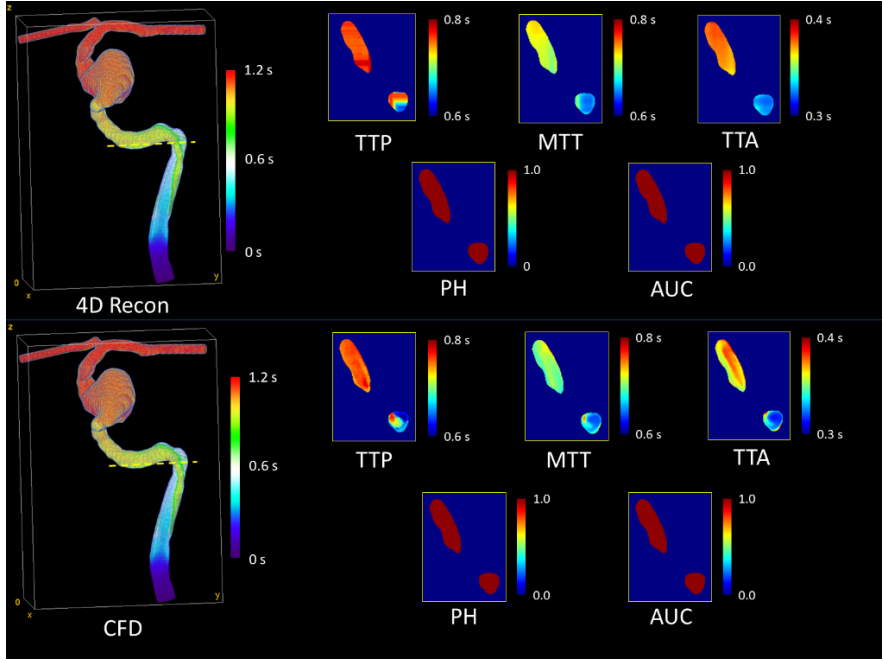

**Figure 5:** Cross-sectional comparison of API results within the ICA (at the yellow dotted line) of model M1 at 35 cm/s inlet velocity. PH and AUC were constant in this cross section due to full filling with contrast media.

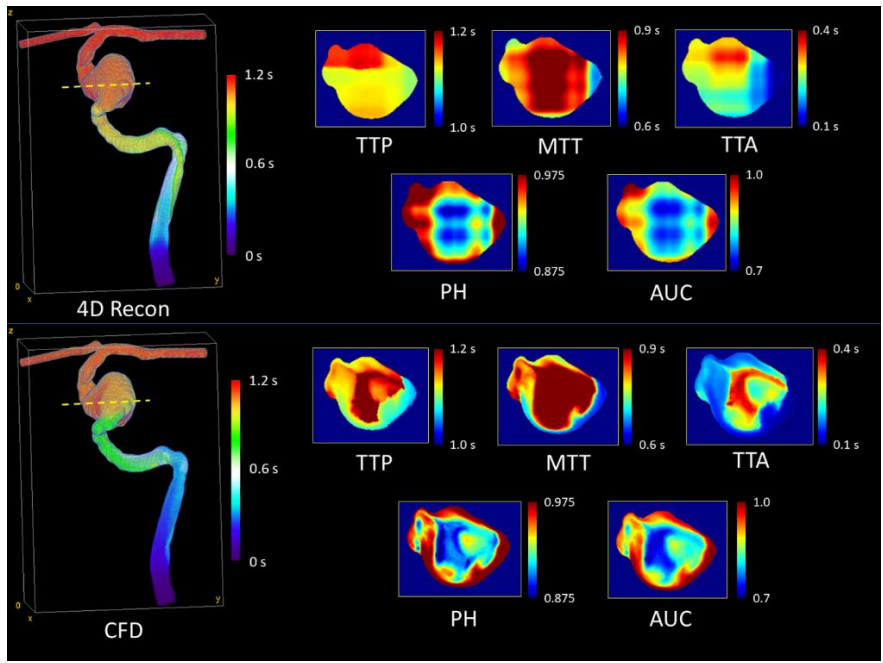

**Figure 6:** Cross-sectional comparison of API results within the aneurysm dome (at the yellow dotted line) of model M1 at 35 cm/s inlet velocity. We observe better agreement between intensity-based parameters, with mild spatial disagreement between datasets across the cross-section.

**Table 2:** Quantitative comparison between API results generated from 4D reconstructed angiograms and ground truth CFD-simulated angiograms at varying blood velocities, including mean absolute error (MAE), mean square error (MSE), and mean absolute percent error (MAPE). For reference, the average magnitude of each parameter in the ground truth API maps is reported (GT Mean).

| | | API Results: 4D Reconstruction vs Ground Truth Simulated Angiograpy | | | | |
|---|---|---|---|---|---|---|
| Model | Velocity (cm/s) | API Parameter | MAE | MSE | MAPE | GT Mean |
| M1 | 25 | TTP | 0.094 | 0.026 | 9.224 | 0.029 |
| | | MTT | 0.047 | 0.005 | 5.014 | 0.023 |
| | | TTA | 0.056 | 0.014 | 14.367 | 0.007 |
| | | PH | 0.047 | 0.006 | 6.458 | 0.024 |
| | | AUC | 0.07 | 0.013 | 9.66 | 0.023 |
| | 35 | TTP | 0.086 | 0.022 | 9.471 | 0.027 |
| | | MTT | 0.043 | 0.004 | 4.963 | 0.022 |
| | | TTA | 0.04 | 0.008 | 14.367 | 0.005 |
| | | PH | 0.032 | 0.003 | 3.633 | 0.025 |
| | | AUC | 0.052 | 0.007 | 7.608 | 0.024 |
| | 45 | TTP | 0.067 | 0.015 | 8.672 | 0.024 |
| | | MTT | 0.035 | 0.003 | 4.43 | 0.02 |
| | | TTA | 0.027 | 0.003 | 13.157 | 0.004 |
| | | PH | 0.011 | 0.001 | 1.136 | 0.027 |
| | | AUC | 0.023 | 0.002 | 2.602 | 0.026 |
| M2 | 25 | TTP | 0.076 | 0.048 | 2.398 | 0.022 |
| | | MTT | 0.023 | 0.001 | 3.055 | 0.016 |
| | | TTA | 0.018 | 0.003 | 10.021 | 0.004 |
| | | PH | 0.001 | 0 | 0.072 | 0.021 |
| | | AUC | 0.006 | 0 | 0.64 | 0.02 |
| | 35 | TTP | 0.051 | 0.025 | 2.014 | 0.016 |
| | | MTT | 0.018 | 0.001 | 3.117 | 0.012 |
| | | TTA | 0.013 | 0.001 | 9.783 | 0.003 |
| | | PH | 0 | 0 | 0.034 | 0.021 |
| | | AUC | 0 | 0 | 0.016 | 0.015 |
| | 45 | TTP | 0.138 | 0.072 | 8.349 | 0.023 |
| | | MTT | 0.011 | 0 | 1.469 | 0.016 |
| | | TTA | 0.011 | 0.003 | 9.521 | 0.003 |
| | | PH | 0 | 0 | 0 | 0.021 |
| | | AUC | 0.003 | 0 | 0.286 | 0.025 |
| M3 | 25 | TTP | 0.073 | 0.029 | 47.427 | 0.025 |
| | | MTT | 0.023 | 0.001 | 3.02 | 0.019 |
| | | TTA | 0.014 | 0.001 | 7.526 | 0.005 |
| | | PH | 0.004 | 0.001 | 0.434 | 0.025 |
| | | AUC | 0.004 | 0.001 | 0.42 | 0.025 |
| | 35 | TTP | 0.051 | 0.015 | 4.667 | 0.018 |
| | | MTT | 0.014 | 0.001 | 2.529 | 0.014 |
| | | TTA | 0.009 | 0 | 7.246 | 0.004 |
| | | PH | 0.004 | 0.001 | 0.401 | 0.025 |
| | | AUC | 0.003 | 0.001 | 0.411 | 0.019 |
| | 45 | TTP | 0.049 | 0.014 | 5.118 | 0.017 |
| | | MTT | 0.011 | 0 | 2.047 | 0.017 |
| | | TTA | 0.007 | 0.001 | 7.148 | 0.003 |
| | | PH | 0.004 | 0.001 | 0.371 | 0.025 |
| | | AUC | 0.003 | 0.001 | 0.386 | 0.019 |

## 4 Discussion

We demonstrate that automatic alignment of biplane angiography (Sec. 2.2) can be applied to 4D reconstruction, provided a 3D vascular mask is also correlated with each view, to generate 4D contrast flow profiles with high temporal resolution. It should be noted that the proposed automated alignment protocol is dependent on temporal synchrony between biplane imagers, as well as an adherence to the assumption that both image systems share a common axial direction (i.e. no cranial or caudal angulation). Fortunately, with our high temporal resolution biplane data, asynchrony between views is quite minimal, meaning the proposed method is valid. At much lower frame rates, less automated techniques of alignment may still be utilized, such as the use of a calibration object[24, 25], or manual delineation of correspondent points between views.[26-29]

The proposed constrained back-projection approach was also quite successful, with an average MSE of 0.007 across the nine reconstructed 4D angiograms relative to ground truth. This correct localization of contrast density values, as a function of both space and time, was also facilitated by the temporal synchrony between our angiographic acquisitions. Although the accuracy of contrast density localization may be affected by the frame rate of each biplane imager, lower frame rates were not explored as part of this study. The largest sources of error between 4D reconstructions and CFD data were mild foreshortening effects as contrast entered vascular structures which were perpendicular to one biplane view, and sections of the vasculature with heterogeneous filling of contrast (Fig. 8), however, these errors were still quite minimal.

The API analysis confirmed the success of our algorithm at temporally resolving volumetric contrast media flow from our biplane data. Intensity-based API parameters were nearly identical between the datasets, which confirms previous reports regarding the benefits of path-length correction as a pre-processing step for projection-based imaging.[15] Disagreements in temporal API parameters were caused by similar factors as were discussed above; vessel foreshortening and incomplete vascular filling can artificially modulate contrast density values as a function of time. For threshold-based, single-timepoint parameters, such as TTP and TTA, this can alter the point at which contrast density is first detected, or where its peak lies, indicated by the relatively elevated error in these parameters relative to others tested in this study. This indicates that other TDC-based parameters which are more robust to these variations may be desired for future 4D reconstructed angiography-based API analyses. This said, previous studies have indicated that MTT, PH and AUC are the three most predictive API parameters for prognosis of aneurysm occlusion,[3, 30] all of which were accurately represented via 4D reconstruction, evidenced by the errors reported in Table 1. Additionally, the strong correlation between API parameters generated from 4D reconstructions and CFD data across many different blood velocities indicates their utility in assessing vascular flow conditions across longitudinal studies. This is duly indicative of the utility of these 4D reconstructions in the assessment of treatment efficacy of endovascular neurointerventions, where API analysis across longitudinal studies has demonstrated prognostic ability.[31] We speculate that this technique could be extended to use in more elaborate, gradient-based quantitative angiography, where 2D DSA is inadequate to fully capture the 3D hemodynamic phenomena within neurovascular pathologies of interest, to provide detailed 3D velocity mapping.[32-36]

This study utilized CFD-generated angiographic data primarily for its convenience both in providing a ground truth 4D dataset to compare to our 4D angiographic reconstructions. There are factors of real angiographic images which must be considered before applying this technique to such data. First, our simulated angiograms are free of quantum mottle and do not factor crosstalk scatter that may be present in the system due to the high frame rate of both image systems.

Although this scatter should not influence the alignment and registration steps used to correlate both biplane views to a common 3D vessel mask, the added variance will influence the constrained back-projection, where quantum mottle effects will be populated across the projection axis of both biplane imagers.

Finally, this feasibility study used orthogonal biplane data only. In the case non-orthogonal views are utilized, the localization of contrast intensities may be diminished slightly, except in the case where a perfectly orthogonal view is along a projection axis with severe vessel foreshortening. This is because, as the angle between imagers diverges from orthogonality, the mutual information in both images increases. Future studies should explore the relationship between biplane angulation and the ability to effectively localize contrast, though this was beyond the scope of this work.

Beyond its immediate applications, this work paves the way for a transformative shift in neurovascular imaging, with potentially far-reaching implications for personalized medicine. The ability to assess neurovascular disease hemodynamic related imaging biomarkers with greater precision could lead to improved risk stratification, better patient selection for intervention, and real-time treatment adaptation. Future studies should focus on validating this approach in vitro and in clinical settings, especially regarding its integration with AI-driven predictive models.[30, 31, 37, 38]

## 5   Conclusions

This study presents a novel reconstruction technique to temporally resolve 3D contrast media flow profiles from biplane, angiographic data, generating a 4D high resolution angiography dataset. This data qualitatively provides insight regarding 3D flow phenomena, and can be used qualitatively for 3D API analysis, suggesting potential prognostic capabilities of the method in the future. The technique could prove useful in neurointerventional suites, both to reduce the need for repeat CTA protocols, and to reduce the need for patient transfer between imaging suites. Implementation of the method could improve visualization and quantitation of the effects of endovascular treatments, further improving patient specificity and standard of care in the clinical setting.

*Disclosures*

The authors of this manuscript have no financial interests to disclose.

*Acknowledgements*

This work was supported by NSF 2304388 STTR Phase II form, QAS.AI Inc and NIH Grant No. 1R01EB030092.

**Kyle Williams** is a research assistant at the SUNY University at Buffalo (Department of Biomedical Engineering). He received his BS degree in biomedical engineering from the SUNY University at Buffalo in 2020 and is pursuing a PhD in biomedical engineering and a Masters in Medical Physics.  His current research interests include imaging informatics, computed tomography reconstruction, machine learning, and computer aided diagnostics.

Biographies and photographs for the other authors are not available.

**Caption List**

**Fig. 1** Constrained back-projection of the same axial slice at three different time points. Each row represents a different time point. The red line in the AP view of the original angiographic

acquisition (A) is used to reference a specific slice of the 4D reconstruction (B) at the indicated time.

**Fig. 2** A sample time-density curve, with each API parameter denoted, including time-to-peak (TTP), mean transit time (MTT), time-to-arrival (TTA), peak height (PH) and area under the curve (AUC). Temporal parameters are shown in red, while intensity-based parameters are shown in blue.

**Fig. 3** Representative time points from each model's reconstructed 4D angiogram are shown across each row. To demonstrate the volumetric nature of the angiograms, the time series is viewed from various angles throughout the sequence.

**Fig. 4** API maps for model M1 at 25 cm/s inlet velocity, generated from the constrained back-projection-reconstructed 4D angiogram. Maps include time-to-arrival (TTA), mean transit time (MTT), time-to-peak (TTP), peak height (PH), and area under the time-density curve (AUC). Temporal parameters are in seconds, while intensity-based parameters are unitless.

**Fig. 5** Cross-sectional comparison of API results within the ICA (at the yellow dotted line) of model M1 at 35 cm/s inlet velocity. PH and AUC were constant in this cross section due to full filling with contrast media.

**Fig. 6** Cross-sectional comparison of API results within the aneurysm dome (at the yellow dotted line) of model M1 at 35 cm/s inlet velocity. We observe better agreement between intensity-based parameters, with mild spatial disagreement between datasets across the cross-section.

**Table 1** Quantitative comparison between reconstructed 4D angiograms and ground truth CFD simulations. Each MSE measurement represents the error across all time steps for a specific inlet velocity.

**Table 2** Quantitative comparison between API results generated from 4D reconstructed angiograms and ground truth CFD-simulated angiograms at varying blood velocities, including mean absolute error (MAE), mean square error (MSE), and mean absolute percent error (MAPE). For reference, the average magnitude of each parameter in the ground truth API maps is reported (GT Mean).